\documentclass{article}
\usepackage{epsf}

\def\Real{{\rm I\mathchoice{\kern-0.70mm}{\kern-0.70mm}{\kern-0.65mm}%
  {\kern-0.50mm}R}}
\def\N{{\rm I\mathchoice{\kern-0.70mm}{\kern-0.70mm}{\kern-0.65mm}%
  {\kern-0.50mm}N}}
\def\C{\rm C\kern-.42em\vrule width.03em height.58em depth-.02em
       \kern.4em}

\newcommand{\be}{\begin{equation}}
\newcommand{\ee}{\end{equation}}
\newcommand{\tb}{$\tau_{\rm burn}\,$}
\newcommand{\Mi}{${\cal M}_i\,$}

\title{Numerical Investigation of Scaling Properties of Turbulent
Premixed Flames}

\author{Jens C. Niemeyer\\ Max-Planck-Institut f\"ur Astrophysik\\
Karl-Schwarzschild-Str. 1\\ 85740 Garching, Germany\\
 \\
and\\
 \\
Alan R. Kerstein\\ Combustion Research Facility\\ Sandia National
Laboratories \\ Livermore, CA 94551-0969, USA}
\begin{document}
\maketitle
\def\baselinestretch{1.3}
\abstract{
Gibson scaling and related properties of flame-surface geometry in 
turbulent premixed combustion are demonstrated using a novel 
computational model, Deterministic Turbulent Mixing (DTM).  In DTM, 
turbulent advection is represented by a sequence of maps 
applied to the computational domain.  The structure of the mapping 
sequence incorporates pertinent scaling properties of the turbulent 
cascade.  Here, combustion in Kolmogorov turbulence (kinetic-energy 
cascade) and in Bolgiano-Obukhov convective turbulence (potential-energy 
cascade) is simulated.  Implications with regard to chemical flames 
and astrophysical (thermonuclear) flames are noted.
}
\clearpage
\section{INTRODUCTION}
\label{intro}

A widely recognized paradigm of turbulent premixed combustion 
is the propagation of a dynamically passive, advected surface at fixed 
speed $u_0$ relative to the fluid.  Despite the omission of important 
flame-turbulence interactions such as flame-generated turbulence and 
stretch-induced variation of the laminar flame speed $u_0$, scaling 
properties deduced from this formulation have proven useful for the 
interpretation of turbulent combustion processes.

Based on the passive-surface picture, three interrelated scaling 
properties that describe the structure and evolution of turbulent 
premixed flames (here considering only the 
flamelet combustion regime) have been proposed, and the predictions 
have been compared to measurements.  First, a scale-invariant law 
governing the dependence of $u_T/u_0$ on $u'/u_0$, where $u_T$ is the 
turbulent burning velocity and $u'$ is the root-mean-square turbulent 
velocity fluctuation, has been derived and confirmed experimentally 
(Pocheau and Queiros-Cond\'e, 1996).  In particular, this law conforms 
to the dimensionally mandated scaling $u_T \sim u'$ for $u'\gg u_0$.  
Second, measurements of the flame-surface fractal dimension (North and 
Santavicca, 1990) are consistent with the scaling prediction (Kerstein, 
1988) $D=7/3$.  Third, it has been proposed that flame-surface 
fluctuations are suppressed below a length scale $l_{\rm g}$ at which 
the generation of fluctuations by turbulence and smoothing of 
fluctuations by flame propagation are in balance.  The balance scale 
$l_{\rm g}$ is determined by setting the eddy velocity equal to $u_0$.  
The inertial-range scaling 
\be
\label{kolmo}
\tau (l) \sim \left(l\over L\right)^{2/3}{L\over u'}
\ee
of the turnover time $\tau (l)$ of size-$l$ eddies is invoked, where $L$ 
is the turbulence integral scale and $u'$ is the rms velocity 
fluctuation.  Taking $l/\tau (l)$, to be the size-$l$ eddy velocity, 
$l_{\rm g}$ is determined from  $l_{\rm g}/\tau(l_{\rm g})=u_0$, yielding 
the Gibson scaling (Peters, 1988) 
\be
\label{gibs}
l_{\rm g}/L \sim (u_0/u')^3.
\ee
Measured flame structure 
does not obey this scaling.  Rather, the available data suggests 
that lower-cutoff behavior is sensitive to flame-turbulence 
interactions omitted from the passive-surface picture (North and 
Santavicca, 1990; G\"ulder and Smallwood, 1995; 
Erard {\it et al.}, 1996).

These three scalings are interrelated in that any two of them 
imply the third, as demonstrated in Sec.~\ref{skol}.  In this regard, the 
failure of Gibson scaling despite the confirmation of the other scalings 
has not been explained.  The approach taken here to address this 
conundrum is to construct an idealized numerical model of dynamically 
passive, propagating interfaces advected by turbulence, formulated so 
that all three scalings are necessarily obeyed in the limit 
$u'\gg u_0$.  The model is used to investigate the transition to 
the asymptotic behaviors as $u'/u_0$ increases.  The numerical 
results suggest an interpretation of the experimental results in 
terms of the influence of flame-turbulence interactions and 
finite-$u'/u_0$ effects.

The discussion thus far is specific to the kinetic-energy cascade in 
incompressible, constant-density turbulence, governed by the Kolmogorov 
scaling, Eq. (\ref{kolmo}).  This is the cascade generated by large-scale 
shear applied, e.g., at flow boundaries.  In buoyant turbulent flows 
with unstable stratification, the large-scale energy input is 
potential energy that is converted to kinetic energy.  Potential 
energy as well as kinetic energy can cascade to smaller scales, and 
exchange between potential and kinetic energy can occur at each scale.  
It has been proposed (L'vov and Falkovich, 1992) that the 
potential-energy cascade may dominate.  This implies the scaling 
\be
\label{bo}
\tau (l) \sim \left( l\over L\right)^{2/5}{L\over u'},
\ee
consistent with an earlier proposal by Bolgiano and Obukhov in a different 
context.  This `Bolgiano-Obukhov' (BO) scaling is controversial; Borue and 
Orszag (1997) discuss experimental, computational, and analytical results 
supporting and contradicting Eq. (\ref{bo}).

Recognizing the considerable if not conclusive support for Eq. (\ref{bo}), 
numerical results are presented for flame propagation in cascades governed 
by Eq. (\ref{bo}).  The inner-cutoff scale and the fractal dimension of 
flames in BO turbulence are predicted from scaling and compared to 
numerical results.  Though this regime is of some relevance to chemical 
combustion, the principal motivation is thermonuclear combustion in 
supernovae.  Implications in this regard are noted.

\section{METHOD}
\label{method}

\subsection{General Technique}
A new computational model, Deterministic Turbulent Mixing (DTM), is 
formulated for the study of multiscale phenomena of interfaces propagating 
in stirred media. Its spirit is to achieve maximum spatial resolution by a 
radical
simplification of specific physical mixing processes, e.g., fully developed
hydrodynamic turbulence. In particular, mixing is characterized here only
by a universal mapping procedure for every length scale and by the
relative frequency of mapping events. Since interface propagation can 
be implemented as a discrete binary process, a method that
fulfills the above description is conveniently based on a cellular
grid of discrete binary states, henceforth termed `unburned' and `burned' 
in analogy with turbulent flame propagation. Furthermore, we have the
freedom to choose the spectral distribution of mixing events directly
and can therefore restrict the calculation to two spatial dimensions
without being constrained by inverse cascade effects of two-dimensional
turbulence.  The method is also applicable in three dimensions, but this 
is not implemented here.

The computational domain is a two-dimensional $(N\times N)$ square
grid of cells of size $(\Delta x)^2$, where $N = 2^r$. Every computation 
starts with the lower half of the domain `burned' and the upper half 
`unburned'.  The algorithm of DTM can then be sketched as follows: during each
burning time step \tb, every neighboring fuel cell of a previously burned
cell is burned, mimicking laminar interface propagation with a speed
$u_0 = \Delta x/$\tb. Here, horizontal, vertical, and diagonal neighbors 
are included. For convenience, we assign $\Delta x =$ \tb = 1
and therefore $u_0 = 1$. Note that only the relative frequency of
mixing and burning events is dynamically
relevant, i.e., the normalization of time and length scales is arbitrary.

Mixing is implemented as a series of discrete events which map the entire
grid onto itself. Each map \Mi convolutes the interface on a
characteristic length scale $l_i=2^i$, where $i_{\rm min}\le i\le 
i_{\rm max}<r$.  A sequence of maps involving $l_i$ values spanning a 
large dynamic range leads to a highly complex, multiscale 
interface structure.  A time scale $\tau_i = \tau(l_i)$ regulates the 
frequency of mapping events on the scale $l_i$ relative to
the interface propagation time step \tb. This function relating the
time and length scales of the problem is the only link to
hydrodynamic turbulence; for instance, by choosing a power law
dependence of $\tau_i$ on $l_i$ we can model different scaling regimes
for the turbulent velocities. The goal of DTM is to study 
interface properties subject to various rules for $\tau_i$ and
rather generic forms for \Mi. In this approach, complexity arises
from a purely deterministic multiscale mixing process coupled to local 
interface propagation. No random elements are involved.

Owing to the absence of floating-point operations, the computational
expense for all the described steps is relatively low. Limited mainly by
memory considerations, a large dynamic range can be achieved and the
evolution can easily be followed over many large scale dynamical times
$\tau_{i_{\rm max}}$. Identification of the largest mixing scale
$l_{\rm max} = 2^{i_{\rm max}}$ with the integral scale $L$ and the
minimum scale $l_{\rm min} = 2^{i_{\rm min}}$ with the Kolmogorov
dissipation scale $l_{\rm k}$ yields an estimate for the effective
turbulent Reynolds number of the Kolmogorov-cascade computations,
\be
\label{re}
Re = \left(\frac{L}{l_{\rm k}}\right)^{4/3} = 2^{4\,\Delta i/3}\,\,,
\ee
with $\Delta i = i_{\rm max} - i_{\rm min}$. 
Calculations reported here involve $\Delta i = 9$, corresponding to
$Re \approx 4100$ for the Kolmogorov cascade. 

On the other hand, this brief description also shows the main
restrictions of DTM. First 
of all, local interface propagation is completely independent of the
mixing process. Strain effects on $u_0$ are omitted. Furthermore, 
feedback of `burning' onto the \Mi and $\tau_i$ is not represented.  
In particular, there is no flame-generated turbulence.  Most 
importantly, the maps \Mi are purely artificial and are not claimed 
to reproduce detailed mixing dynamics of hydrodynamic turbulence. 
They can be expected, however, to belong to the same universality class 
as a large ensemble of scale invariant turbulent eddies, and thus 
provide useful insight into the statistical behavior of stirred 
propagating interfaces.
 
\subsection{Implementation of Mixing}
Given the general philosophy of DTM as described above, the following
mixing properties can be varied according to the physical system under
consideration: the form of the maps \Mi = ${\cal M}(l_i)$, the mapping
frequencies $\tau_i^{-1} = \tau^{-1}(l_i)$ (equivalent to choosing a
power spectrum of a stochastic mixing process), and their
normalization with respect to the local interface propagation speed
$u_0$. 

We shall first describe the procedure for mixing on a single
length scale $l_i$ for a given frequency $\tau_i^{-1}$. The choice of
\Mi is the least constrained from first principles. Features that 
should be incorporated into
discrete maps in reduced turbulence models have been summarized
(Kerstein, 1991). By construction, their effect is intended to be a
combined stretching and folding of the interface on the length scale $l_i$. 
Consequently, they contain the single parameter $l_i$, and their
dependence on the spatial coordinates is preferably chosen to be
continuous and differentiable in order to reflect smooth physical
mixing.

As a specific representation for \Mi, we choose a sinusoidal deformation 
of the entire grid, analogous to the action of a single Fourier
mode. For a given length scale $l_i$, the sequence of maps 
alternates between deformation in the $x$ and $y$ directions.  The 
time between successive maps for given $l_i$ is $\tau_i$.  Then for 
given $l_i$, the mapping sequence is:
\be
\label{mi}
{\cal M}_i(n): \{x,y\} \to 
\left \{ 
\begin{array}{ll}
\{x,y+{\rm int}[l_i\,\sin(\pi \, x/2 l_i)]\} & {\rm if} \quad t =
2n\,\tau_i \\   
\{x+{\rm int}[l_i\,\sin(\pi \, y/2 l_i)],y\} & {\rm if} \quad t =
(2n+1)\,\tau_i  
\end{array}\right.
\ee
for $n = 0,1,\dots$, where $t$ denotes the epochs at which the
indicated maps are implemented and int[.] stands for the nearest integer.
Periodic boundary conditions are imposed horizontally, and we assume
completely burned or unburned material at the bottom and top boundaries,
respectively.  The computational domain is shifted vertically as needed 
to prevent burned material from contacting the top boundary.

Figure 1 demonstrates the effect of the alternating $x$ and
$y$-maps, Eq. (\ref{mi}), for a single length scale $l_i = r-3$ over a time 
period  of 6 $\tau_i$.  Here, burning is not implemented.  Starting with a 
planar interface, the \Mi deform it step by step into rolls resembling 
vortices of a Kelvin-Helmholtz (KH) unstable shear layer. Maps acting on 
multiple length scales can thus be envisioned to produce a superposition of 
KH vortices on all scales.

This picture is specific to the mathematical definition, 
Eq. (\ref{mi}), of the mapping sequence.  In particular, 
Eq. (\ref{mi}) is formulated so that the point $(0,0)$ and its periodic 
images never move.  If successive mappings are allowed to have different 
sets of fixed points, more complicated patterns can be obtained for 
fixed $l_i$.  The intent here is to obtain the simplest possible behavior 
for fixed $l_i$ so that multiscale structure mainly reflects the 
assumed distribution of $l_i$.  For purposes such as the study 
of flame surface response to local turbulence structure, some variant of 
this formulation might be preferable.  A three-dimensional formulation 
can readily be configured to generate helical fluid paths, reminiscent of 
flow near vortex tubes.  Any such analogy between the model and properties 
of turbulent flow would require detailed validation that has not yet been 
undertaken.  The intent here is to formulate a model that emulates 
particular scaling properties of propagating surfaces advected by 
turbulence, irrespective of possible broader implications of the model.

Generalization to multiple length scales is achieved by concurrent 
implementation of the map sequences ${\cal M}_i(n)$ consecutively for all 
$i \in [i_{\rm min}, i_{\rm max}]$.  If events corresponding to more than 
one scale $i$ occur at a given epoch, then the events are implemented in 
order of increasing $i$. (Smaller-scale events are governed by shorter 
time scales and therefore are deemed to be `completed' before concurrent 
larger-scale events.  Scaling properties are insensitive to this somewhat 
arbitrary interpretation.)  The dependence of the event frequency on $l_i$ 
is derived from the notion of a velocity spectrum $v_i = v(l_i)$ of 
turbulent fluctuations in scale space. We define 
\be
\tau_i = \frac{l_i}{v_i}
\ee
in analogy with the turnover time of size-$l_i$ eddies. By
postulating a power law for the velocity distribution in scale space,
\be
\label{vl}
v_i \propto \left( l_i \over \lambda \right)^\eta V\,\,,
\ee
where $\lambda$ and $V$ are reference length and velocity scales 
(analogous to the turbulence integral scale and rms velocity fluctuation), 
we can link the inverse mapping frequencies (the time between two
consecutive mapping events) directly to the length scales: 
\be
\label{ti}
\tau_i \propto \left(l_i \over \lambda \right)^{1-\eta}{\lambda \over V}\,\,.
\ee
Unless otherwise noted, we will henceforth set $\eta = 1/3$, relating
the DTM mixing process specifically to inertial-range Kolmogorov
turbulence.

In order to normalize the mixing frequencies with respect to the
interface propagation, the parameter
\be
\label{z}
z = \frac{u_0 \tau(l_{\rm k})}{l_{\rm k}}
\ee 
is introduced.  As in Eq. (\ref{re}), the smallest mixing scale 
$l_{i_{\rm min}}$ is interpreted as the dissipation scale $l_{\rm k}$.  
In the derivation of Eq. (\ref{gibs}) it was assumed that 
inertial-range scaling applies at the scale $l_{\rm g}$, and thus that 
$l_{\rm g}>l_{\rm k}$.  This is equivalent to the requirement $z>1$.  
In all of our calculations, $z \ge 2$.

$z$ parameterizes the small-scale mixing.  The large-scale mixing is 
parameterized by $u'/u_0$, which is defined in the DTM context as 
$u'/u_0=L/[u_0\tau (L)]$, where the largest mixing scale $l_{i_{\rm max}}$ 
is interpreted as the turbulence integral scale $L$.  This definition omits 
the contribution of mixing at scales below $L$.  A more literal analog of 
the turbulent velocity fluctuation $u'$ can be defined within DTM.  The 
choice of definition rescales the numerical results, but does not affect 
the properties of interest here.

Based on Eq. (\ref{re}), these parameters obey the relation 
$zu'/u_0 =Re^{1/4}$ for the Kolmogorov cascade.  The turbulent combustion 
process can thus be parameterized by $u'/u_0$ and either $Re$ 
(the usual choice) or $z$.  $z$ is convenient for identifying the range of 
validity of Gibson scaling, and therefore is adopted here.  For all 
Kolmogorov-cascade computations reported here, $Re=2^{12}$, so $u'/u_0=8/z$.

Common parameters of all computations (except an illustrative case shown 
in Figure 2) were
the grid size, $N = 2^{12}$, and the indices $i_{\rm min} = 1$ and 
$i_{\rm max}= 10$ determining the scales $l_{\rm k} = 2^{i_{\rm min}}$ 
and $L = 2^{i_{\rm max}}$ of the smallest and largest maps.  $z$
ranged from 2 to 4, yielding $2 < u'/u_0 < 4$ for the Kolmogorov cascade.  
For the Bolgiano-Obukhov cascade, this $z$ range corresponds to 
$10.5< u'/u_0 < 21$.

Figure 2 shows a typical flame image from a computation based on the 
Kolmogorov scaling.  The multiscale nature of the flame interface is 
apparent.

\section{RESULTS AND INTERPRETATIONS}
\subsection{Kolmogorov Inertial-Range Turbulence}
\label{skol}
Here, DTM is used to investigate the interface properties that can be 
quantified in terms of the box-counting statistics described 
by Menon and Kerstein (1992).  This is implemented by partitioning the 
flame image into square boxes with side lengths $l_j = 2^j,
\,j = 0 \dots r-2$ and counting the number $b(l_j)$ of boxes that
contain at least one newly burned cell. Obviously, $b(l_0)$ is the total 
number of newly burned cells, so $u_T/u_0$ is the ensemble average of 
$b(l_0)/N$, where $b(l_0)/N$ may be interpreted geometrically as the 
interface arc length per unit transverse length.  Accordingly, 
$A_j=l_jb(l_j)/(l_0N)$ may be interpreted as the arc length per unit 
transverse length of the smoothed interface, for smoothing scale 
$l_j$.  Henceforth, $A_j$ is denoted $A(l)$, omitting the subscript on 
$l$.

If $l$ is much smaller than the finest scale of surface wrinkling, then 
the smoothing has no effect, so $A(l)$ should converge to a constant 
value in the limit of small $l$.  The flattening of computed profiles of 
this quantity at small $l$ is seen in Figure 3.

For $l$ much larger than the flame-brush thickness, the smoothed interface 
is planar, so $A(l)$ converges to unity.  The computational domain is 
not large enough to demonstrate this asymptote, but this is immaterial 
because the unity asymptote is a mathematical necessity.  More 
significant is the tendency for the profiles to collapse with respect to 
$l/L$ with increasing $l$.  This indicates that the flame-brush thickness 
scales as $L$.

Interpreting the two-dimensional computation as a planar cut through 
three-dimensional space, fractal scaling implies $A(l)\sim l^{2-D}$, 
where $D$ is the fractal dimension of the flame surface in 
three-dimensional space (North and Santavicca, 1990).  This scaling 
is identified by linear dependence on a log-log plot of $A(l)$.

The profiles in Figure 3 suggest linear scaling between the inner and 
outer cutoffs.  The linear range appears to broaden with increasing 
$u'/u_0$ (corresponding to decreasing $z$), as one would expect.  
The slope increases with $u'/u_0$, and in fact, 
the implied fractal dimensions fall within the error bars on a plot of 
measured $D$ as a function of $u'/u_0$ (North and Santavicca, 1990).  The 
predicted value $D=7/3$ is not attained by the computed profiles, consistent 
with the experimental finding that this occurs only for $u'/u_0$ much higher 
than the computed cases (North and Santavicca, 1990).

Next, it is shown that any two of the three scalings discussed in the 
Introduction imply the third.  Consider a box-counting profile, plotted in 
the format of Figure 3, for $u'/u_0\gg 1$.  For $l\ll l_{\rm g}$, it 
approaches a plateau level of order $u'/u_0$ because $u_T \sim u'$.  
For $l\gg L$, it approaches a plateau level of unity.  Implicit 
here is the assumption that the flame-brush thickness is of order $L$.  
However, this is not an additional assumption.  Rather, this assumption 
and $u_T \sim u'$ both follow from the assumption that flame-brush 
length and time scales are governed by large-eddy length and time scales 
for $u'\gg u_0$.

Thus, the ratio of the small-$l$ and large-$l$ levels is $u'/u_0$.  
In Figure 3, this vertical span is traversed over a horizontal span 
$L/l_{\rm g}$.  If there is fractal scaling, then the profile has a region 
of linear dependence between these limits.  In the limit $u'\gg u_0$, 
the slope $2-D$ of this linear region, based on these estimates, is 
$-\log (u'/u_0) /\log (L/l_{\rm g})$.  This gives 
$l_{\rm g}/L \sim (u_0/u')^{1/(D-2)}$, showing that the fractal scaling 
implies the Gibson scaling and vice versa.  Analogous reasoning shows 
that if both scalings are obeyed, then $u_T\sim u'$.

The apparent implication of the computed results in Figure 3 and of 
the measurements is that fractal scaling is more robust than Gibson scaling, 
because fractal scaling is obtained, but for $D$ values inconsistent with 
Gibson scaling.  However, a different picture is obtained when $Au_0/u'$ is 
plotted with respect a Gibson-normalized horizontal axis, as in Figure 
4.  Here, $l_{\rm g}$ is defined as $(u_0/u')^3L$.  
In this format, Gibson scaling, in conjunction with $u_T\sim u'$, implies 
collapse of profiles below the outer cutoff.  $u_T\sim u'$ by itself implies 
only a collapse of the low-$l$ plateau levels.  The profiles indicate some 
deviation from the latter collapse, suggesting either deviation from 
this scaling in the $u'/u_0$ range of the computations, or numerical 
resolution effects.  Resolution effects can be discounted only if the 
computations resolve the low-$l$ plateau over a significant $l$ range.  
It is evident that the resolution is marginal by this criterion.

Whatever the source of low-$l$ deviation, the collapse between the 
inner and outer cutoffs indicates accurate conformance to Gibson 
scaling.  It is also evident that the universal scaling function thereby 
obtained has significant curvature throughout the range in which it is 
well resolved.  Although an eventual transition to fractal scaling with 
$D=7/3$ is a necessary consequence of the convergence of the profiles 
to a universal form (and the assumption that the outer cutoff scales as 
$L$), it appears that the transition from the low-$l$ plateau to the 
fractal scaling extends over more than three decades of $l$.

Thus, Figure 4 indicates that both the Gibson scaling and the relation 
$u_T\sim u'$ are more robust than fractal scaling because they are both 
seen at $u'/u_0$ for which the latter is not seen.  How can this be 
reconciled with Figure 3, and with measurements?

The apparent linear regions in Figure 3 may be manifestations of the 
inflection points of the profiles, which necessarily occur owing to profile 
flattening at small and large $l$.  Measured profiles used to infer fractal 
dimensions typically have no greater range of apparent linearity, oftentimes 
less.  Comparison of Figures 3 and 4 shows that 
extraction of fractal dimensions from such data is problematic.  A more 
reliable method for determining scaling properties is to demonstrate that 
a particular scaling collapses a family of curves parameterized by 
$u'/u_0$.

Onset of fractal scaling may be slow because the mathematical requirement 
of spatial homogeneity is not strictly obeyed.  Flame geometry varies with 
streamwise location.  For example, isolated pockets of unburned material 
are more likely to be found on the upstream side of the flame brush (e.g., 
Figure 2).  Therefore, flame geometry is approximately fractal only when 
the flame interface near the flame-brush center becomes so wrinkled that 
it dominates the box-occupancy statistics over a significant dynamic 
range.  In this case, local homogeneity near the flame-brush center is 
sufficient for approximate fractal behavior.  This condition can be 
fulfilled only for high $u'/u_0$.

In contrast, Gibson scaling reflects the local balance between wrinkling 
by advection and smoothing by propagation.  Because DTM has been formulated 
so that propagation and advection are both spatially homogeneous, this 
scaling should not vary with streamwise location.  This expectation is 
supported by the apparent robustness of Gibson scaling.  However, in 
combustion processes with flame-turbulence interactions such as 
flame-generated turbulence, turbulence intensity will, in general, vary 
with streamwise location relative to the flame brush, so the onset of 
Gibson scaling may be inhibited.

\subsection{Convective Turbulence}
\label{sbo}

DTM allows freedom of choice for the assumed turbulent velocity spectrum. 
The impact of a particular cascade mechanism on turbulent flame propagation 
is investigated by fixing the parameter $\eta$ in equation (\ref{vl}) to the 
relevant scaling exponent. We used $\eta = 1/3$ in the 
calculations summarized in Sec.~\ref{skol}.

A particular example of an eddy cascade whose velocity
scaling exponent differs from $\eta = 1/3$ is buoyancy-driven turbulence. 
As indicated in Sec.~\ref{intro}, the coupled density and velocity
cascades are presumed to be governed by the Bolgiano-Obukhov (BO) scaling 
$\eta = 3/5$. The scaling properties of premixed flames propagating in 
buoyancy-driven flows are derived on this basis.  Following the arguments 
of Kerstein (1988), we 
expect a fractal dimension of $D_{\rm BO} = 2\frac{3}{5}$. The lower 
cutoff scale, denoted $l_{\rm w}$, is the balance scale at which the 
eddy velocity $(l/L)^{\eta}u'$ is of order $u_0$, giving 
\be
\label{lw}
l_{\rm w} = (u_0/u')^{5/3}L\,\,.
\ee

DTM provides a simple illustration of the above relations.
The results of computations with $\eta = 3/5$ are shown in Figure
5.  They involved $z = 2$, 3, and 4, corresponding to $u'/u_0=
21$, 14, and 10.5. 
The format of Figure 4 is used, except that $l$ is 
normalized by $l_{\rm w}$ in order to demonstrate the BO analog of Gibson 
scaling.  Collapse of the computed profiles verifies this scaling.  As in 
Figure 4, it is seen that there is a gradual transition to 
the anticipated fractal scaling.

These results have significant implications
in the context of astrophysics. A specific class of extremely bright
astronomical events, so-called Type Ia supernovae, can be consistently
explained by thermonuclear explosions of compact white dwarf stars 
(Niemeyer and Woosley, 1997, and references therein). In this scenario, 
the thermonuclear reaction wave propagates from the stellar core outwards 
as a microscopically thin flame sheet, `burning' carbon and oxygen to
iron-group elements. The resulting stratification of hot, burned
material underneath the dense outer layers is Rayleigh-Taylor
unstable, rapidly giving rise to the production of strong
turbulence. Hence, the situation fulfills all of the criteria for
flame propagation in buoyancy-driven turbulence and, consequently, our
results obtained for the BO cascade apply. This question is subject to
ongoing research; a more detailed analysis will be presented
elsewhere (Niemeyer and Kerstein, 1997). 

\section{CONCLUSIONS}
\label{con}

Previous work has shown that analysis of a dynamically passive, 
propagating surface advected by turbulence is a useful framework 
for interpreting turbulent combustion phenomena.  Here, the assumptions 
generally adopted in this framework have been embodied in a 
computational model, Deterministic Turbulent Mixing (DTM).

The model provides an explicit numerical demonstration of the scalings 
governing turbulent premixed flame structure and burning velocity in 
the flamelet regime.  Moreover this demonstration has several 
implications with regard to turbulent combustion measurements.  The 
main results and conclusions are as follows.

First, the Gibson scaling of the inner cutoff of flame structure in 
Kolmogorov turbulence is demonstrated by obtaining the collapse of 
a family of box-occupancy profiles, parameterized by $u'/u_0$, with 
appropriate normalization.  This is the first explicit demonstration 
that Gibson scaling of a propagating surface follows directly from 
Kolmogorov scaling of turbulent advection.

Second, the structure of the 
universal scaling function revealed by this collapse is compared to 
the structure of individual profiles.  Though individual profiles 
suggest fractal scaling with fractal dimension $D$ that increases 
with $u'/u_0$, the universal function shows no fractal scaling over 
the scale range probed by the computation.  In effect, the different 
$D$ values of the individual curves reflect the fact that each 
individual curve corresponds to a slightly different range, in the 
Gibson-normalized coordinate, of the universal curve.  Thus, the 
apparent $u'/u_0$ dependence of $D$ is a measure of the curvature 
of the universal function.

Third, the results have bearing on the interpretation of turbulent 
combustion measurements despite the simplifications inherent in 
the model.  The results indicate that the measured $u'/u_0$ 
dependence of $D$ may in fact be an indication of lack of fractal 
scaling, rather than variation of a scaling exponent.  Measurements 
suggest eventual convergence to $D=7/3$, but convincing convergence 
is not attained over a significant dynamic range of $u'/u_0$.  Thus, 
one possible explanation of the empirical failure of Gibson scaling 
is that $u'/u_0$ large enough to observe the scaling has not yet been 
achieved in the experiments.  However, model results suggest that 
Gibson scaling, in the passive-surface framework, is more robust 
than fractal scaling.  This suggests that a more likely explanation 
of Gibson-scaling failure is flame-turbulence interactions omitted 
from the passive-surface formulation.  If this is true, then the 
convergence to $D=7/3$, if it occurs, may reflect a combination 
of flame-turbulence interaction effects rather than applicability of 
the scaling analysis.

Fourth, the model demonstrates the proposed connection between 
turbulence structure and flame structure.  When the Bolgiano-Obukhov 
scaling presumed to govern convective turbulence is adopted in lieu 
of the Kolmogorov scaling, the inner-cutoff scaling changes 
accordingly.  The analysis implies a corresponding change of $D$, 
but this cannot be demonstrated within the computationally accessible 
range of scales.  It is interesting to note that the passive-surface 
modeling framework may be more relevant to buoyancy-driven thermonuclear 
flames than to chemical flames because the astrophysical flames 
have relatively low thermal expansion, hence weak flame-turbulence 
interactions (except for gravitational forcing), and because $u'/u_0$ 
in these flames is `astronomical.'

\section{ACKNOWLEDGMENTS}

The authors would like to thank R. Rinne for facilitating this 
collaboration and S. Woosley for consenting to the use of his initial.  
This research was partially supported by the Division of Engineering 
and Geosciences, Office of Basic Energy Sciences, U.S. Department of 
Energy. The computations were performed at the Rechenzentrum Garching, 
Germany. 

\par\vfill\eject

\clearpage
\begin{figure}
\epsfxsize = 16cm
\vspace{-7cm}
\epsfbox{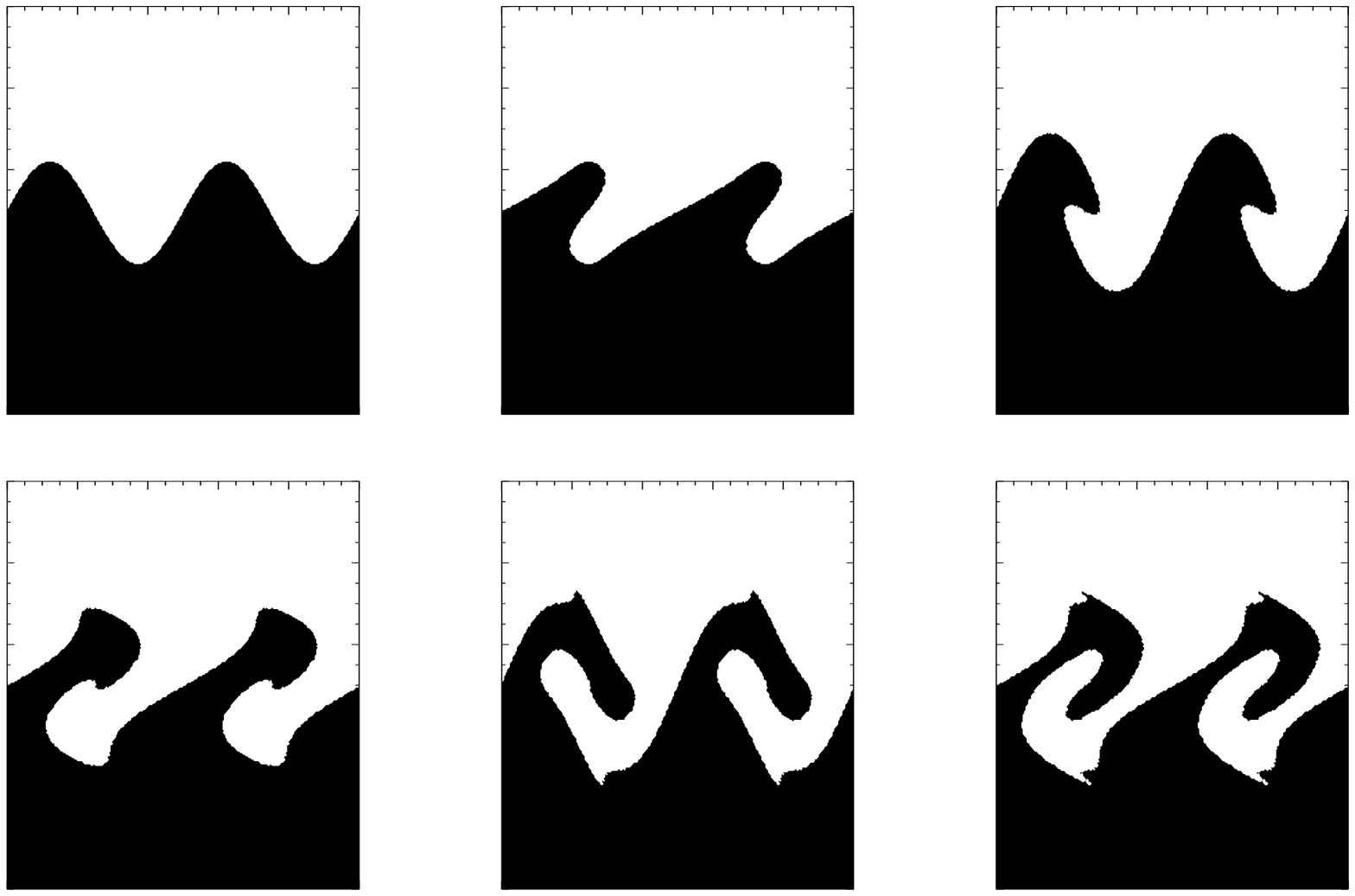}
\vspace{-3cm}
\caption{Snapshots of the interface during the first 6
iterations of the alternating mapping sequence, Eq. (\ref{mi}). For this
demonstration, only a single length scale, $l_i = 32$, was mapped and
interface propagation was turned off. The resolution was $N = 2^8$.  
The sense of rotation can be reversed by changing $+$ to $-$ in either the 
top or the bottom line of Eq. (\ref{mi}}
\end{figure}

\begin{figure}
\epsfxsize = 16cm
\vspace{-7cm}
\epsfbox{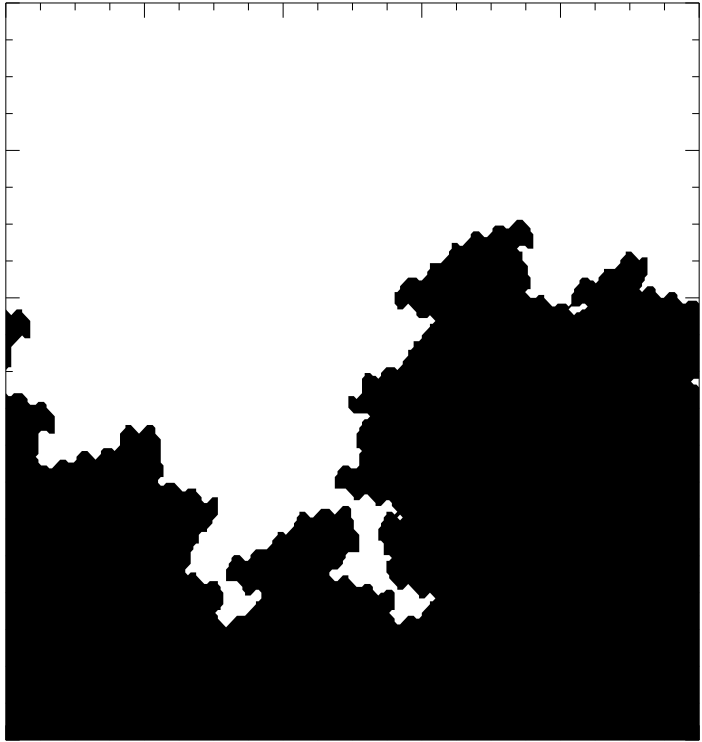}
\vspace{-3cm}
\caption{Image of a flame in a simulated Kolmogorov cascade after 
$\approx 30$ large scale mapping times. Black is burned.  The computation 
used $N = 2^8$, $i_{\rm min} = 1$, $i_{\rm max} = 6$, and $z = 1$.}
\end{figure}

\begin{figure}
\epsfxsize = 16cm
\vspace{-7cm}
\epsfbox{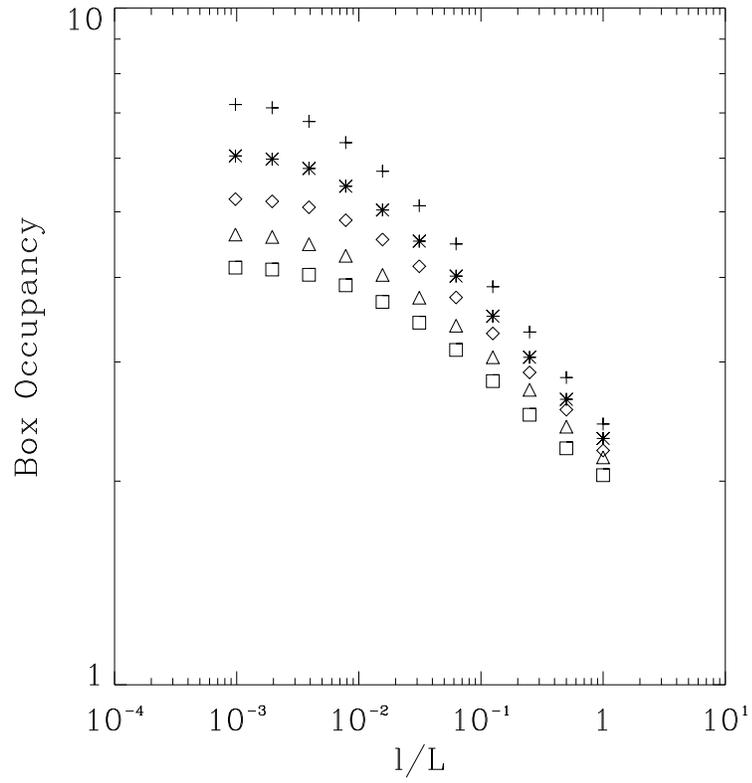}
\vspace{-3cm}
\caption{Normalized box occupancy $A$ as a function of length 
scale, normalized to the integral scale $L$. The symbols represent, from 
top to bottom: $z = 2, 2.5, 3, 3.5, 4$.}
\end{figure}

\begin{figure}
\epsfxsize = 16cm
\vspace{-7cm}
\epsfbox{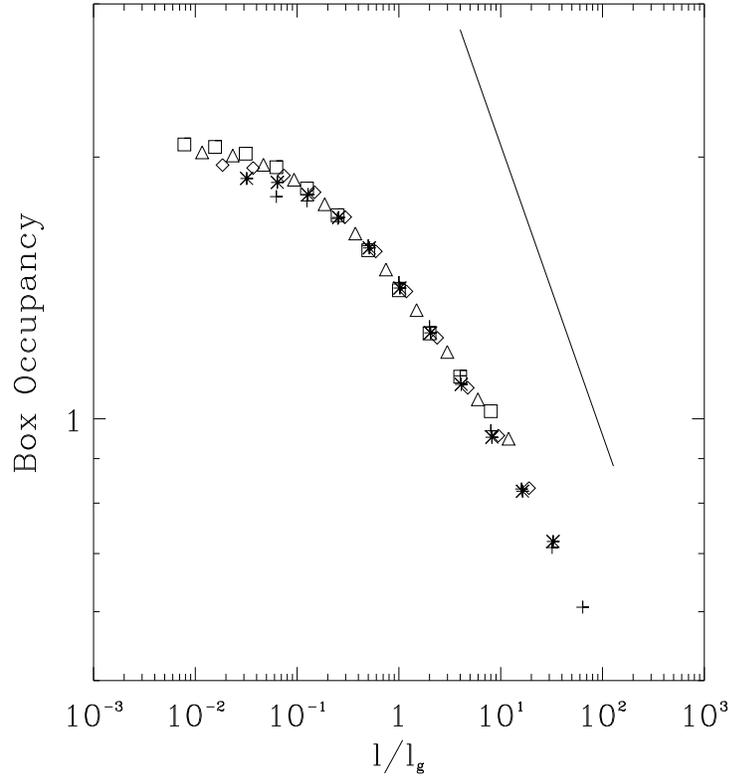}
\vspace{-3cm}
\caption{Rescaled box occupancy $Au_0/u'$ as a function of length scale, 
normalized to the Gibson scale $l_{\rm g}$.  The correspondence between 
symbols and $z$ values is as in figure 3.  The line corresponds to a linear 
decline with a slope of $-1/3$.}
\end{figure}

\begin{figure}
\epsfxsize = 16cm
\vspace{-7cm}
\epsfbox{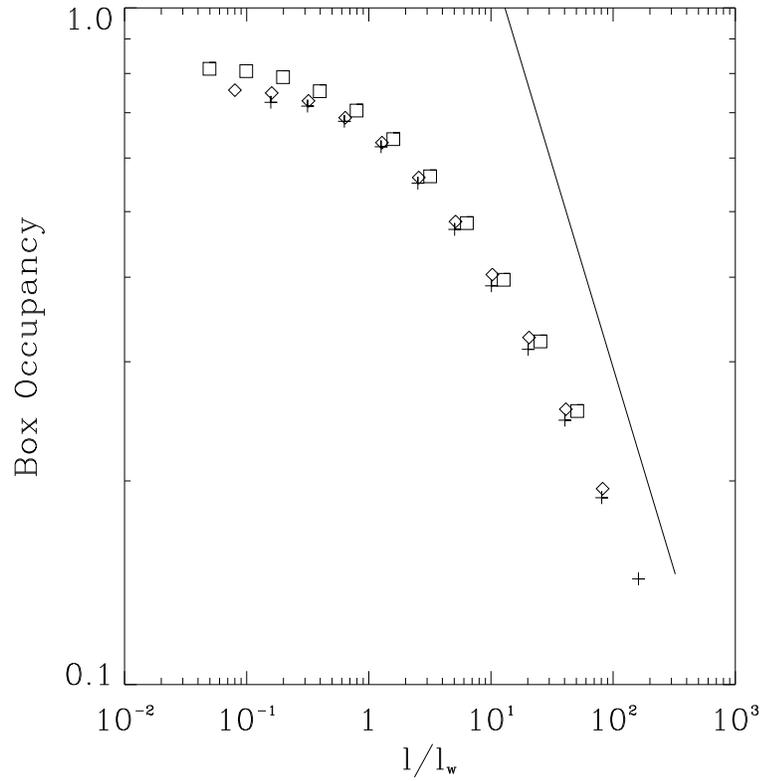}
\vspace{-3cm}
\caption{Rescaled box occupancy $Au_0/u'$ in convective turbulence 
(eddy-velocity scaling exponent $\eta =5/3$) as a function of length scale, 
normalized to the cutoff scale $l_{\rm w}$.  The correspondence between 
symbols and $z$ values is as in figure 3.  The line corresponds to a 
linear decline with a slope of $-3/5$.}
\end{figure}

\end{document}